\documentclass[aps,prb,showpacs,twocolumn]{revtex4-1}
\usepackage{amsmath,amssymb,epsfig,graphicx}
\begin{document}
\title{Nonmagnetic and ferromagnetic fcc cerium studied with one-electron methods}
\author{Fabien Tran}
\author{Ferenc Karsai}
\author{Peter Blaha}
\affiliation{Institute of Materials Chemistry, Vienna University of Technology,
Getreidemarkt 9/165-TC, A-1060 Vienna, Austria}

\begin{abstract}

Density functional theory was used to study the nonmagnetic (NM) and
ferromagnetic (FM) phases of face-centered cubic cerium. Functionals of four
levels of approximation for the exchange-correlation energy were used:
LDA, PBE, LDA/PBE+$U$, and YS-PBEh. The latter two contain an adjustable
parameter, the onsite Coulomb repulsion parameter $U$ for LDA/PBE+$U$ and the
fraction $\alpha_{\text{x}}$ of Hartree-Fock exchange for YS-PBEh, which were
varied in order to study their influence on the results.
By supposing that, as a first approximation, the NM and FM solutions can be
identified to the observed $\alpha$ and $\gamma$ phases, respectively,
it is concluded that while a small value of $U$ or $\alpha_{\text{x}}$ leads to
the correct trend for the stability ordering of the two phases,
larger values are necessary for a more appropriate (but still not satisfying)
description of the electronic structure.

\end{abstract}

\maketitle

\section{Introduction}

Cerium shows an isostructural (fcc $\rightarrow$ fcc) pressure-induced phase
transition associated with a large 15\% change of volume at room temperature.
\cite{KoskenmakiHPCRE78} In the large-volume $\gamma$ phase, which is accessible
above $\sim200$ K, a Curie-Weiss behavior for the magnetic susceptibility is
observed. Application of pressure drives cerium into the small-volume
$\alpha$ phase, which shows Pauli paramagnetism.
Many experimental and theoretical studies have been conducted in order
to understand the mechanism underlying this phase transition. Essentially two
models have been proposed. In the Mott transition model \cite{JohanssonPM74} the
$4f$-electrons undergo a transition from a localized nonbonding character (in
the $\gamma$ phase) to an itinerant bonding character (in the
$\alpha$ phase), while the $spd$ electrons are not considered to play
any significant role. The other proposed mechanism is the Kondo volume
collapse model \cite{AllenPRL82} in which hybridization between the $4f$ and
$spd$ electrons is taken into account and leads to a screening of the local
$4f$ moment which is stronger in the $\alpha$ phase than in the $\gamma$ phase.
Photoemission and bremsstrahlung isochromat spectroscopy experiments
\cite{WuilloudPRB83,WieliczkaPRB84,WeschkePRB91} showed that upper and lower
Hubbard bands are present (i.e., the $4f$ electrons are strongly
correlated) in both phases, while a quasiparticle peak is observed only in the
$\alpha$ phase, indicating a reduced strength of correlation in the $\alpha$ phase.

From a theoretical point of view, the approaches that have been used to study
the $\alpha$ and/or the $\gamma$ phases include
the local density (LDA) and generalized gradient (GGA) approximations,
\cite{PickettPRB81,JohanssonPRL95,SoderlindPRB95,JarlborgPRB97,JarlborgPRB98,CasadeiPRL12}
the self-interaction corrected LDA (SIC-LDA) method,
\cite{SzotekPRL94,SvanePRL94,SvanePRB96,LaegsgaardPRB99,LudersPRB05}
LDA/GGA+$U$,
\cite{ShickJESRP01,AmadonPRB08,WangPRB08,AmadonJPCM12}
LDA plus orbital polarization (LDA+OP),
\cite{ErikssonPRB90}
the LDA plus Gutzwiller approximation (LDA+GA),
\cite{TianPRB11,LanataPRL13}
LDA plus dynamic mean-field theory (LDA+DMFT),
\cite{ZolflPRL01,HeldPRL01,McMahanPRB03,HaulePRL05,McMahanPRB05,SakaiJPSJ05,RueffPRL06,AmadonPRL06,PourovskiiPRB07,AmadonJPCM12,LanataPRL13,ChakrabartiPRB14}
$GW$,
\cite{SakumaPRB12}
and a combined hybrid/Hartree-Fock+random-phase approximation (HF+RPA) study
\cite{CasadeiPRL12} (see Ref. \onlinecite{NikolaevPU12} for a summary).

In general, the proper treatment of solids containing strongly correlated electrons
with the Kohn-Sham (KS) equations\cite{KohnPR65} of density functional theory
(DFT)\cite{HohenbergPR64} is not an easy task and,
in particular, the results obtained with LDA or GGA are very often qualitatively
incorrect.\cite{TerakuraPRB84} Therefore, more advanced methods should be used for
such solids, and
most of them combine LDA (or GGA) with other theories (HF, DMFT, etc.).
Since the $4f$ electrons in the $\alpha$ phase of cerium are sometimes believed
to be less localized (and therefore \textit{less strongly correlated}) than
in the $\gamma$ phase (in particular in the Mott picture mentioned above),
then a fair description of the $\alpha$ phase could eventually be obtained with
the semilocal (LDA/GGA) functionals.
On the other hand, for a correct description of the more correlated $\gamma$
phase, a method beyond LDA/GGA, like DFT+$U$, has to be used.

In this work, we will present the results of a detailed DFT study on cerium.
Four different levels of approximation for the exchange-correlation
functional were used for the calculations: LDA, GGA, LDA/GGA+$U$, and GGA-hybrid.
We will focus on the relative stability of the nonmagnetic
(NM) and ferromagnetic (FM) phases of fcc cerium as well as on their
electronic structures.
As done in most of the previous DFT studies cited above, we will suppose that the
NM and FM solutions of our calculations represent the experimentally observed
$\alpha$ and $\gamma$ phases, respectively. However, it is not clear how
legitimate such an identification can be considered. Actually, in the $\alpha$ phase
it is not known exactly to what extent the local magnetic moment is quenched
(the results of recent experiments suggest that the instantaneous moment
remains stable across the transition\cite{LippPRL12}) and a more appropriate
modelization of the paramagnetic $\gamma$ phase with static mean-field methods
(like those used in the present work) should be done with a
supercell containing randomly oriented moments, while all results of the
present work were obtained in the one-atom unit cell. Obviously, the
state-of-the-art method for such paramagnetic systems is LDA+DMFT, which
treats more rigorously correlation effects.
Therefore, this aspect of
the calculations should be kept in mind when comparing our results with
experiment.

In Refs. \onlinecite{AmadonPRL06,DecrempsPRL11} it was concluded that
at low-temperature the $\alpha$ phase should be more stable than the
$\gamma$ phase by 20$-$30 meV, and in Ref. \onlinecite{CasadeiPRL12} it
was shown that the HF+RPA method is able to predict the correct stability
ordering, while the hybrid functionals HSE06\cite{HeydJCP03,KrukauJCP06} and PBE0
\cite{ErnzerhofJCP99,AdamoJCP99} can not. Actually, the correct stability
ordering was previously reproduced with the LDA+OP
(Ref. \onlinecite{ErikssonPRB90}) and LDA+SIC (Ref. \onlinecite{LudersPRB05})
methods, and in Ref. \onlinecite{WangPRB08} it was shown that the GGA+$U$
method with a small value of $U$ also leads to the correct trend.
At non-zero temperature, the LDA+DMFT method produces a
depression in the total-energy curve, which is consistent with the volume collapse
transition (see, e.g., Ref. \onlinecite{HeldPRL01}).
Here, we will show that the LDA/GGA+$U$ and GGA-hybrid methods lead to very
similar results for the relative energy of the NM and FM phases as well as
the electronic structure. In particular, by tunning
the onsite Coulomb parameter or the fraction of HF exchange, the correct
stability ordering can be obtained.
However, none of the one-electron methods considered in this work
(and in previous works), including the GGA-hybrid, is able to reproduce
all features observed in the experimental spectra and
therefore, as already concluded from previous works, it seems that the
many-body effects seen in the spectra are such that they can not be
\textit{mimicked} by a one-electron method.

The present work is an attempt to give an overview of the
(un)suitability of one-electron methods in general (KS and mixed KS/HF) to
reproduce the experimental facts (lattice constant, stability ordering,
and electronic properties) at low temperature. 

The paper is organized as follows. In Sec. \ref{details}, the theoretical
method is briefly outlined and the computational details are given.
In Sec. \ref{results} the results are presented and discussed, and in
Sec. \ref{summary} the summary is given.

\section{\label{details}Computational details}

The calculations were done with the all-electron WIEN2k code,\cite{WIEN2k}
which is based on the full-potential (linearized) augmented plane-wave plus
local orbitals method\cite{Singh} to solve the KS equations.\cite{KohnPR65}
The one-atom fcc unit cell was used for the calculations and the integrations
into the Brillouin zone were done with a $20\times20\times20$ \textbf{k}-mesh
for the LDA, PBE, and LDA/PBE+$U$ functionals and a $12\times12\times12$ \textbf{k}-mesh
for the much more expensive hybrid functional YS-PBEh.\cite{TranPRB11}
$R_{\text{MT}}K_{\text{max}}=9$ (8 for YS-PBEh), the product of
the atomic sphere radius $R_{\text{MT}}^{\text{Ce}}=2.2$ bohr
and the plane wave cutoff parameter $K_{\text{max}}$,
was used for the expansion of the basis set.
An estimation of the error bar in our calculations for the relative energies of
the NM and FM phases is 25 meV for the hybrid functional YS-PBEh and 10 meV for the
other functionals. The symmetry constraint was reduced from cubic to
orthorhombic, which is enough to lift the degeneracies of the $f$-orbitals
and to allow an orbital moment to develop if spin-orbit coupling (SOC) is included.
A symmetry breaking in Ce could also be explained by the fact that since
Ce is paramagnetic, a Ce atom is surrounded by atoms with
randomly oriented magnetic moments, which breaks the cubic symmetry.
As a side remark, we note that if SOC is not included, then
the electronic states NM and FM1 (see Sec. \ref{occupationmatrix}) can be
reached with cubic symmetry, while FM2 and FM3 (for which also
SOC is necessary) can not.
All the presented results, including those on the NM phase, were obtained
without imposing any constraint on the spin polarization.

Several functionals, namely LDA, PBE, LDA+$U$, PBE+$U$, and YS-PBEh, were considered for
the present work. LDA is the exact functional for the uniform electron
gas\cite{KohnPR65} and for the correlation part the analytical form of Perdew
and Wang,\cite{PerdewPRB92} PW92, was used.
The PBE functional of Perdew, Burke, and Ernzerhof
\cite{PerdewPRL96} is of the GGA form and is the most used functional
for solid-state calculations.
The LDA/PBE+$U$ functionals read
\begin{equation}
E_{\text{xc}}^{\text{LDA/PBE}+U} = E_{\text{xc}}^{\text{LDA/PBE}} +
E_{\text{ee}} - E_{\text{dc}},
\label{ExcPBEpU}
\end{equation}
where $E_{\text{ee}}$ is an (rotationally invariant) electron-electron (ee)
interaction energy of HF-type \cite{LiechtensteinPRB95,ShickPRB99} for the
electrons of a selected atom and angular momentum $\ell$ ($\ell=3$ for the
$4f$ electrons of cerium) and $E_{\text{dc}}$ is the double-counting (dc)
term for which we chose the fully localized limit version.
\cite{AnisimovPRB93,CzyzykPRB94,LiechtensteinPRB95} $E_{\text{ee}}$ and
$E_{\text{dc}}$ depend on the occupation matrix $n_{mm'}$. Since in cerium the
number of $4f$ electrons is one, the results depend mainly only on the
difference $U-J$ between the Coulomb parameter $U$ and the exchange parameter
$J$. Therefore, for simplicity we chose to set $J=0$ in all our calculations.

In the screened hybrid functional YS-PBEh,\cite{TranPRB11} (YS stands
for Yukawa screened), a fraction
$\alpha_{\text{x}}$ ($\in [0, 1]$) of
short-range (SR) PBE exchange is replaced by SR-HF exchange [usually, the acronym
(YS)-PBE0 is used when $\alpha_{\text{x}}=0.25$]:
\begin{equation}
E_{\text{xc}}^{\text{YS-PBEh}} = E_{\text{xc}}^{\text{PBE}} +
\alpha_{\text{x}}\left(E_{\text{x}}^{\text{SR-HF}} -
E_{\text{x}}^{\text{SR-PBE}}\right),
\label{ExcYSPBE}
\end{equation}
where $E_{\text{x}}^{\text{SR-HF}}$ and $E_{\text{x}}^{\text{SR-PBE}}$ are obtained
from their unscreened counterparts by replacing the bare Coulomb operator by the
exponentially attenuated Yukawa operator as proposed by Bylander and Kleinman.
\cite{BylanderPRB90}
Note that the YS-PBEh functional leads to results which are
very similar to the screened hybrid HSE06\cite{HeydJCP03,KrukauJCP06}
functional which uses the complementary error function for the screening (see Ref.
\onlinecite{TranPRB11} for more details about YS-PBEh and
Refs. \onlinecite{BotanaPRB12,TranPRB12,LaskowskiPRB13,KollerJPCM13} for
recent applications).
Spin-orbit coupling was included in the calculations using LDA,
PBE, and LDA/PBE+$U$, but not YS-PBEh since at the moment it is not possible
to include SOC in a calculation which uses the HF method.
However, we could see that SOC has a small effect on the equilibrium
lattice constants ($<0.01$ \AA), relative energies ($<10$ meV), and
electronic structure.

\section{\label{results}Results and discussion}

\subsection{\label{occupationmatrix}Occupation matrix}

Before starting the discussion of the results, we would like to point out the
problem of the multiple solutions that can possibly be obtained with
approximate functionals.\cite{WagnerPRL13}
With functionals which lead to an orbital-dependent potential
(e.g., DFT+$U$ or hybrid), multiple solutions can easily be obtained for
systems with open $d$- or $f$-shell, and actually the $d$- or $f$-orbitals
occupation that is obtained at the end of the self-consistent field (SCF)
procedure strongly depends on (and will eventually be more or less the same as)
the orbitals occupation that is used to start the SCF procedure.
\cite{ShickJESRP01,AmadonPRB08} Therefore,
with such functionals it is recommended to start a SCF calculation with each
of the most plausible orbitals occupations in order to check which one leads
to the lowest energy. In the following, we discuss the different states
obtained with the various functionals.

\begin{figure}
\begin{picture}(8,8)(0,0)
\put(0,4){\epsfxsize=4cm \epsfbox{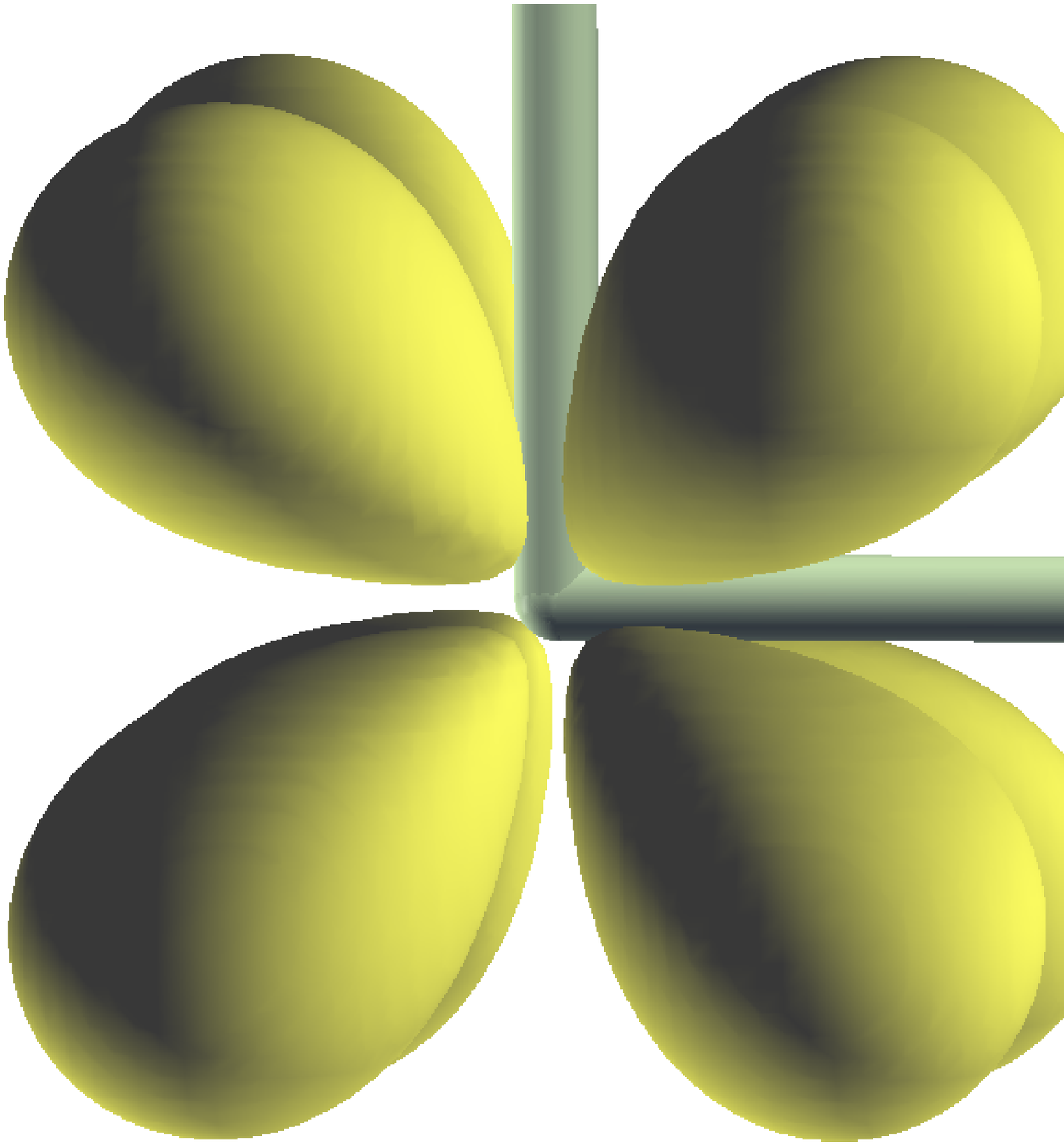}}
\put(4,4){\epsfxsize=4cm \epsfbox{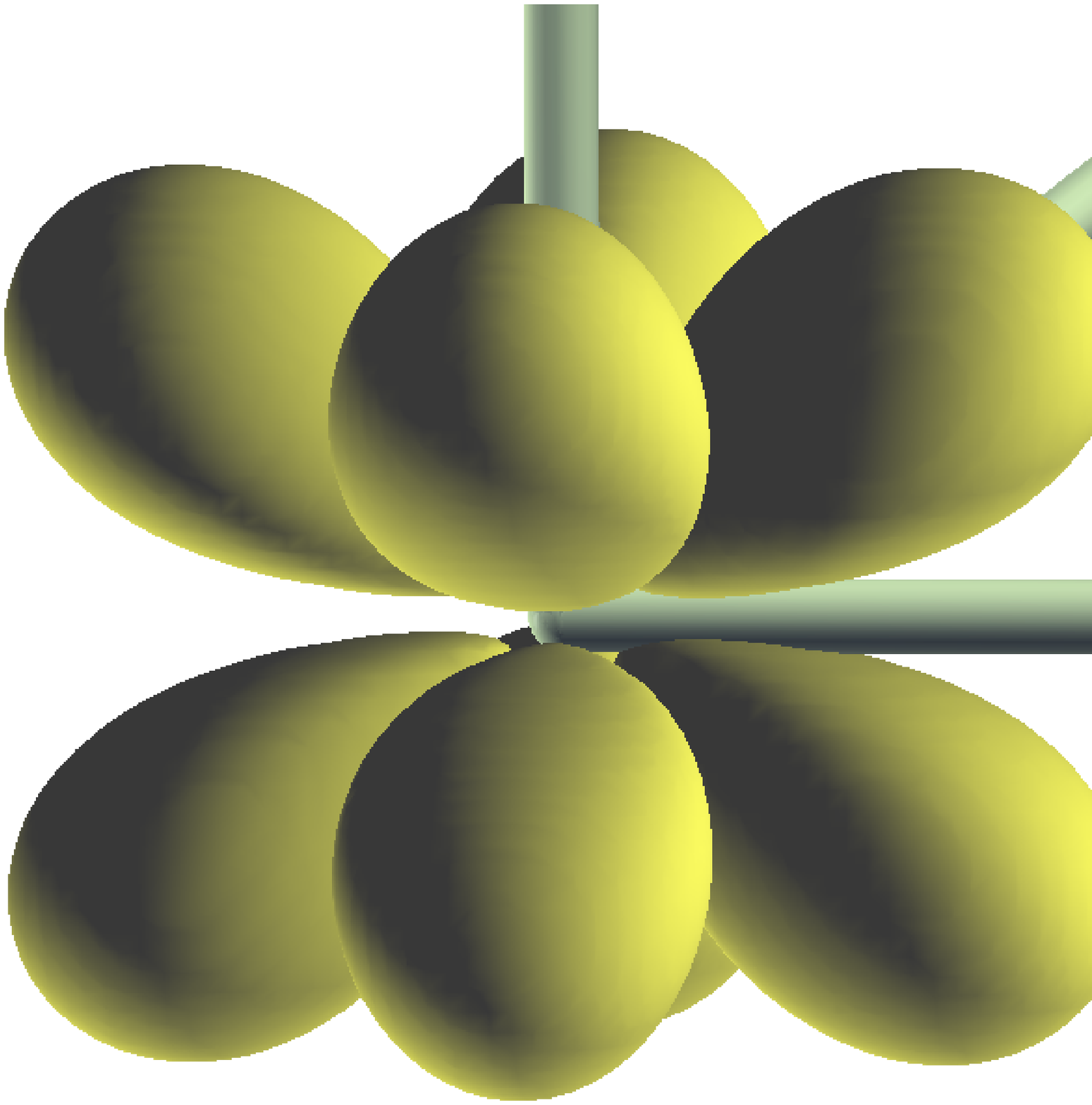}}
\put(0,0){\epsfxsize=4cm \epsfbox{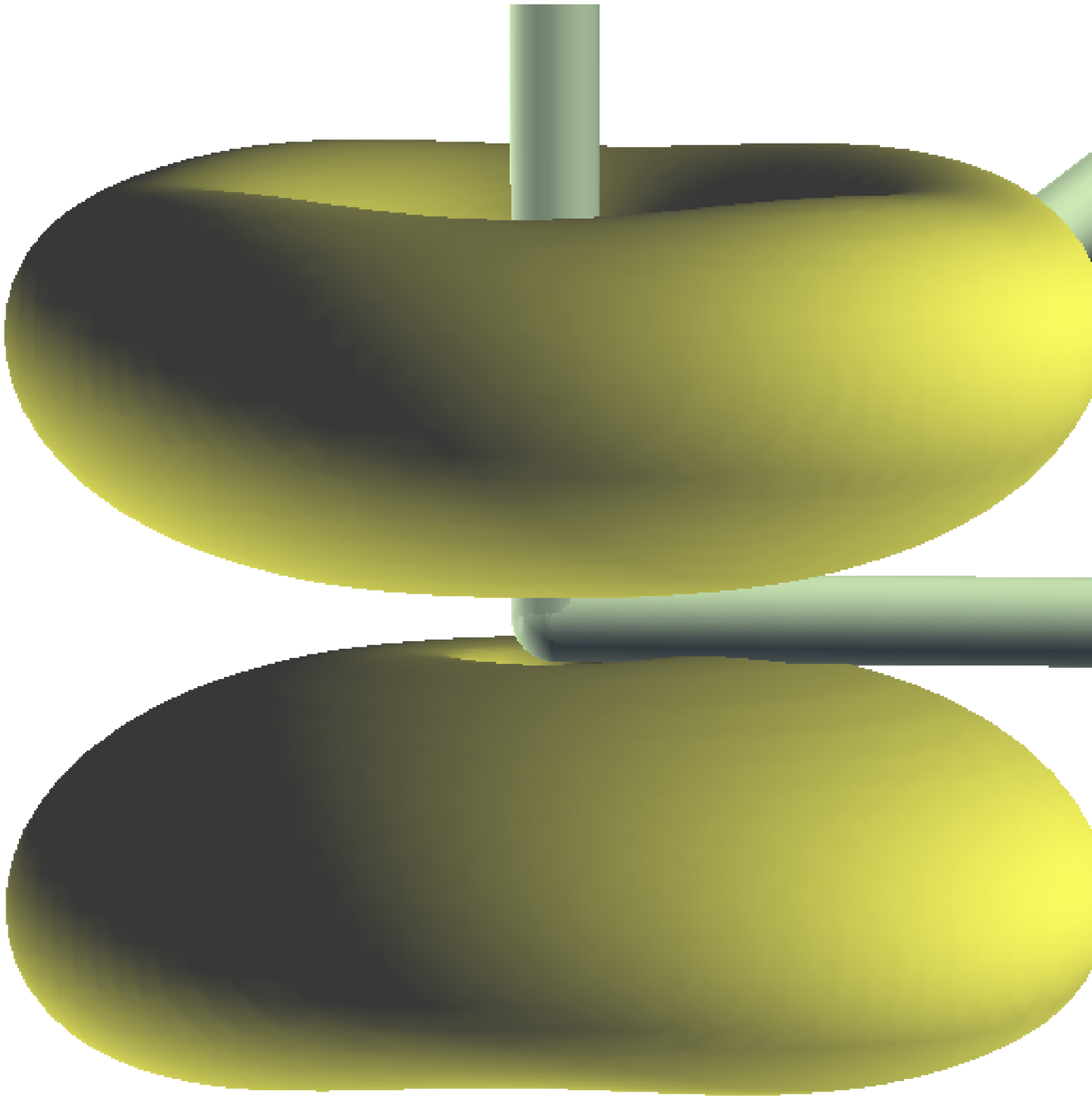}}
\put(4,0){\epsfxsize=4cm \epsfbox{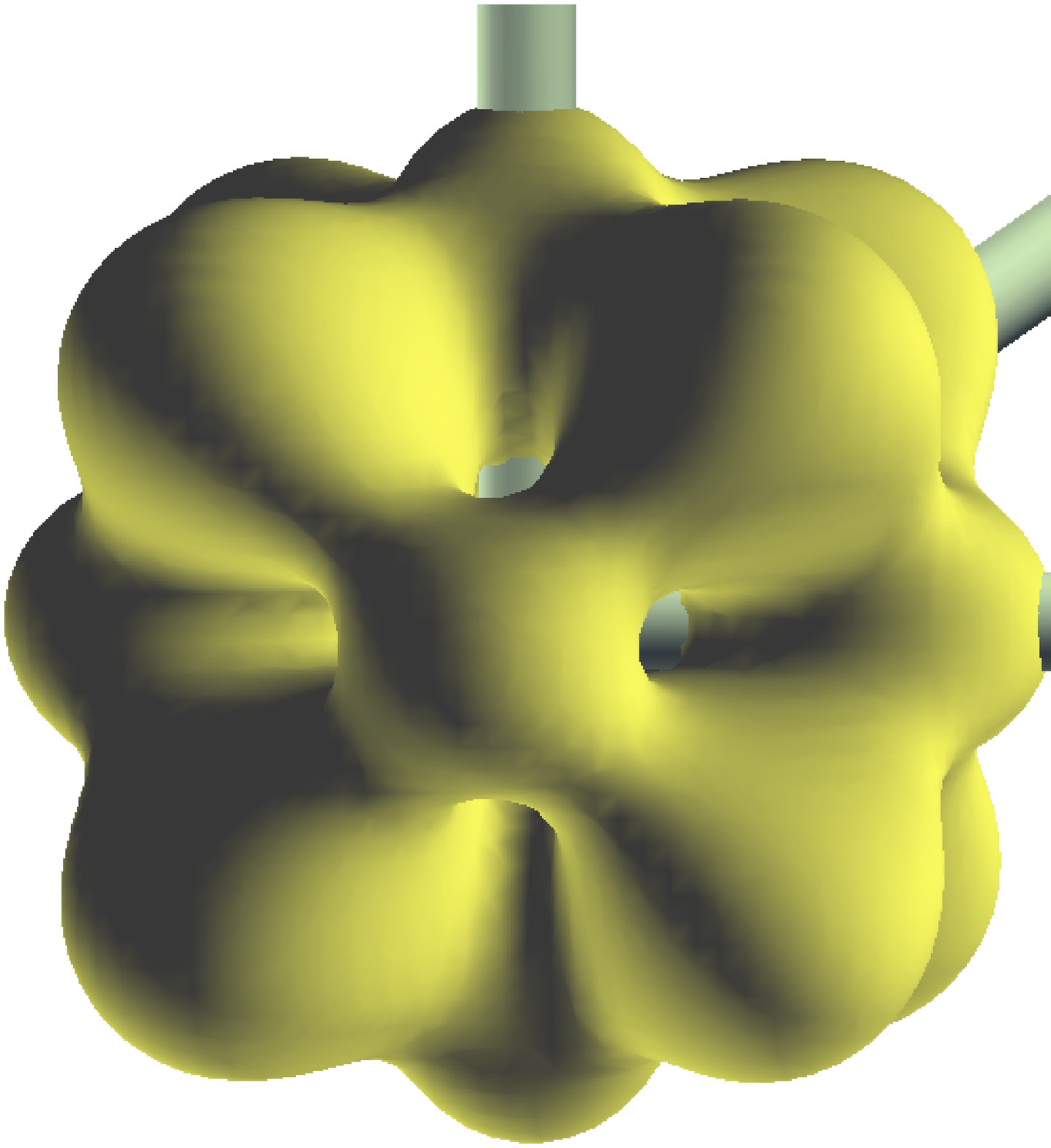}}
\put(0.1,7.8){(a)}
\put(4.1,7.8){(b)}
\put(0.1,3.8){(c)}
\put(4.1,3.8){(d)}
\end{picture}
\caption{\label{fig1}(Color online) The spin-majority $4f$-electron density
at an isovalue of 0.1 electron/bohr$^{3}$ for the FM1 (a), FM2 (b), FM3 (c), and
NM (d) phases obtained from PBE+$U$ with $U=4.3$ eV and SOC. The shown axis are
those of the conventional cubic fcc unit cell.}
\end{figure}

Cerium has (about) one $4f$ electron, and for the FM phase, among the
solutions that we could stabilize with PBE+$U$ (about ten, but probably more
can be stabilized), the two lowest solutions
consist of linear combinations of $Y_{3}^{-2}$ and $Y_{3}^{2}$, which is
in accordance with Refs. \onlinecite{LudersPRB05,AmadonPRB08,CasadeiPRL12}.
The occupation matrices of these two FM solutions, called FM1 and FM2,
are given in the Appendix \ref{occupation} for the case $U=4.3$ eV including SOC.
Note that FM1 and FM2 correspond (approximately since SOC is included)
to an electron in the orbitals
$f_{xyz}=\left(Y_{3}^{2}-Y_{3}^{-2}\right)/\left(i\sqrt{2}\right)$
and $f_{z(x^{2}-y^{2})}=\left(Y_{3}^{2}+Y_{3}^{-2}\right)/\sqrt{2}$, respectively.
It is noteworthy to mention that FM1 is the solution that was obtained by
starting the PBE+$U$ calculations from the PBE electron density,
however, while FM1 is more stable than FM2 for small values of $U$,
it is FM2 which is the most stable for larger values of $U$ (see details below).
In Ref. \onlinecite{ShickJESRP01}, the solution corresponding to an
electron mainly in $Y_{3}^{-2}$, reacheable only if SOC is included,
was found to be the most stable with LDA+$U$ among the states considered
by the authors. It was possible to stabilize a similar state (called FM3)
with PBE+$U$, however as shown in Sec. \ref{latticeconstant} it
is less stable than FM1 and FM2. For the small-volume NM phase, only one
solution could be stabilized and it corresponds to more or less equal occupancies
of the diagonal terms as well as some off-diagonal terms
[see Eq. (\ref{ocNM})]. The occupation matrix of the NM phase is rather
similar among all functionals that we considered (from LDA to YS-PBEh).
The spin-majority $4f$-electron density
is shown in Fig. \ref{fig1}, where we can see that it consists of 14 small
lobes in the NM case [Fig. \ref{fig1}(d)], while there are 8 lobes for FM1
[$f_{xyz}$, Fig. \ref{fig1}(a)] and FM2 [$f_{z(x^{2}-y^{2})}$, Fig. \ref{fig1}(b)]. 

The FM1 and FM2 solutions could apparently not be stabilized with LDA+$U$
when SOC is included, therefore, only NM and FM3 (the solution found in
Ref. \onlinecite{ShickJESRP01} with LDA+$U$) will be considered for this functional.
However, we mention that the FM3 solutions obtained with LDA+$U$ and PBE+$U$
differ slightly in the sense that with PBE+$U$, all elements except $(-2,-2)$
of the occupation matrix are zero [see Eq. (\ref{ocFM3})], while in the
case of LDA+$U$, the off-diagonal terms $(\pm2,\mp2)$ have a value of about 0.2.
Therefore, the character of the FM3 solution obtained with LDA+$U$
is intermediate between the FM2 and FM3 solutions obtained with PBE+$U$.
In a recent study\cite{SchronJPCM13} it has been shown that the gradient
correction in GGA is responsible of the quenching of the orbital magnetic
moment in FeO and CoO. The same mechanism could eventually explain why
the LDA+$U$ most stable FM solution (FM3) corresponds to a large orbital moment,
while it is not the case with PBE+$U$ since the orbital moments are much
smaller in the case of FM1 and FM2 (results discussed in
Sec. \ref{electronicstructure}).

The PBE+$U$ electron densities were used to start the calculations with the
hybrid functional YS-PBEh, and it was found that for two selected
lattice constants, FM1 is more stable than FM2 for all values of
$\alpha_{\text{x}}$ that we considered. Since calculations with the YS-PBEh
functional are expensive, only the FM1 (and NM) solutions will be considered
for YS-PBEh in the following.

The occupation matrix of the FM solution stabilized with LDA and PBE (called
FM in the following) does not differ too much from the NM occupation matrix,
but shows a more pronounced $f_{xyz}$ character.

In Ref. \onlinecite{CasadeiPRL12}, magnetic moments of about 1 and
0.2 $\mu_{\text{B}}$ for the large-volume and small-volume phases,
respectively, were obtained with the PBE0 functional. However, we have not been
able to stabilize such a solution with a small magnetic moment of
0.2 $\mu_{\text{B}}$ (all solutions mentioned above for the FM phase correspond
to a spin magnetic moment of 1.1-1.4 $\mu_{\text{B}}$), neither with PBE+$U$ nor
with YS-PBEh. In particular, the use of the fixed-spin moment method
\cite{SchwarzJPF84} to stabilize such a state has been unsuccessful.

\subsection{\label{latticeconstant}Lattice constant and relative energy}

\begin{figure}
\includegraphics[scale=0.7]{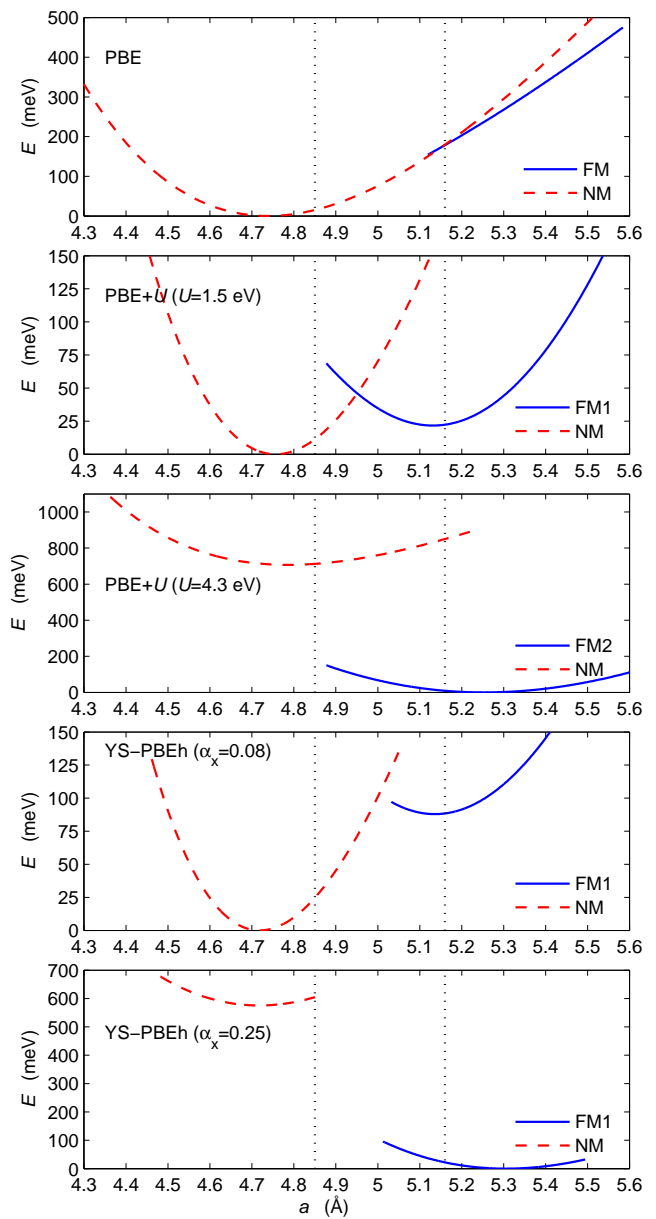}
\caption{\label{fig2}(Color online) Total energy of the FM (blue solid curve)
and NM (red dashed curve) phases of cerium versus the lattice constant calculated
with different functionals.
The vertical dotted lines indicate the experimental values of the $\alpha$
(4.85 \AA) and $\gamma$ (5.16 \AA) phases.
The zero of the energy was set at the minimum of the most stable phase.
Note that the scale on the energy axis is different for each functional.
The results for PBE and PBE+$U$ include SOC effects.}
\end{figure}

The results for the lattice constants and relative energies are shown in
Fig. \ref{fig2} and Table \ref{table1}. As already shown previously,
\cite{SzotekPRL94,JohanssonPRL95} LDA strongly underestimates the lattice constant by
$\sim0.3$ \AA~compared with the experimental value of 4.85 \AA~for the $\alpha$
phase,\cite{KoskenmakiHPCRE78} while the use of a GGA functional such as PBE leads to
better agreement, albeit there is still an underestimation of 0.1 \AA.
With LDA and PBE it was also possible to stabilize a
FM solution, but only for values of the lattice constant $a$ larger than
$\sim5.1$ \AA~as shown in Fig. \ref{fig2}. At the largest value of
the lattice constant that we considered ($\sim5.55$ \AA),
the spin magnetic moment in the unit cell is quite large
($\sim1.4$ $\mu_{\text{B}}$ for LDA and $\sim1.6$ $\mu_{\text{B}}$ for PBE),
but then decreases when $a$ gets smaller to finally disappear when
the FM curve (smoothly) joins the NM curve at about 5.1 \AA.

For the LDA/PBE+$U$ functional, several values of $U$ in the range 0$-$8 eV were
considered, and for two of them the results are shown in
Fig. \ref{fig2} and Table \ref{table1}.
A value of $U=1.5$ eV leads to quite satisfactory results within the
PBE+$U$ method. The minima of the NM and FM1 (more stable than FM2)
curves are at 4.76 and 5.13 \AA,
respectively, the latter value being in good agreement with the
experimental value of 5.16 \AA\cite{KoskenmakiHPCRE78} for the $\gamma$ phase.
More importantly, the NM phase is more stable than the FM1 phase by $\Delta E=-22$ meV,
which is in agreement with the range of values (from $-20$ to $-30$ meV) deduced from
experiment.\cite{AmadonPRL06,DecrempsPRL11} The calculated transition pressure
amounts to $P_{t}=-0.5$ GPa, which seems to be smaller than the experimental value
which should be around $-1$ GPa.\cite{KoskenmakiHPCRE78}
Actually, Wang \textit{et al}. \cite{WangPRB08} already
showed that at $T=0$ K, a value of $U=1.6$ eV (also with PBE+$U$)
leads overall to the most consistent results and in particular to the
correct stability ordering. The FM2 and FM3 solutions are less stable
than FM1 by 18 and 55 meV, respectively, and the lattice constant for
FM3 is shorter than for FM1 and FM2 by $\sim0.1$ \AA.
The results for LDA+$U$ (Table \ref{table1}) show that the lattice constants of the NM and
FM phases are strongly underestimated. We found that for this
functional, a value around $U=2.7$ eV
leads to good agreement with experiment for $\Delta E$.

Recently, in Ref. \onlinecite{NilssonPRB13} the constrained RPA method was
used to calculate the (static) parameters $U$ and $J$ for
the early lanthanides. The values for $U-J$ that were obtained for the
$\alpha$ and $\gamma$ phases are 3.8 and 4.8 eV, respectively
(this difference between the two phases is a consequence of their different
lattice constants). The results obtained with the average (4.3 eV) of these
two values for $U$ in our PBE+$U$ calculations (we recall that we set $J=0$)
are given in Fig. \ref{fig2} and Table \ref{table1}.
In contrast to what was obtained with $U=1.5$ eV, we can see that
with $U=4.3$ eV, the FM solutions are more stable than the NM
one by 600$-$700 meV, which is not the correct stability ordering.
Furthermore, the lattice constants of the
FM phases are now too large by 0.1 \AA, while for the NM phase
there is still a sizable underestimation compared to experiment
(4.79 \AA~with $U=4.3$ eV versus 4.85 \AA~for experiment).

\begin{table*}
\caption{\label{table1}Equilibrium lattice constant $a_{0}$ (in \AA) and
total-energy difference $\Delta E=E^{\text{NM}}-E^{\text{FM}}$ (in meV)
between the minima of the FM and NM phases. A negative value of $\Delta E$
indicates that the NM phase is more stable than the FM phase.
The results for LDA, PBE, LDA+$U$, and PBE+$U$ include SOC effects.}
\begin{ruledtabular}
\begin{tabular}{lccccccc}
Method                             & $a_{0}^{\text{NM}}$ & $a_{0}^{\text{FM1}}$ & $a_{0}^{\text{FM2}}$ & $a_{0}^{\text{FM3}}$ & $\Delta E^{\text{FM1}}$ & $\Delta E^{\text{FM2}}$ & $\Delta E^{\text{FM3}}$ \\
\hline
LDA                                & 4.52 \\
PBE                                & 4.74 \\
LDA+$U$ ($U=2.7$ eV)               & 4.58                &                      &                      & 4.91                 &                         &                         & $-32$                   \\
LDA+$U$ ($U=4.3$ eV)               & 4.60                &                      &                      & 5.00                 &                         &                         & 362                     \\
PBE+$U$ ($U=1.5$ eV)               & 4.76                & 5.13                 & 5.16                 & 5.06                 & $-22$                   & $-40$                   & $-77$                   \\
PBE+$U$ ($U=4.3$ eV)               & 4.79                & 5.26                 & 5.25                 & 5.25                 & 692                     & 707                     & 616                     \\
YS-PBEh ($\alpha_{\text{x}}=0.08$) & 4.72                & 5.14                 &                      &                      & $-88$ \\
YS-PBEh ($\alpha_{\text{x}}=0.25$) & 4.72                & 5.31                 &                      &                      & 576   \\
\cline{3-5}\cline{6-8}
\multicolumn{1}{l}{Expt.} &
\multicolumn{1}{l}{4.85\footnotemark[1]} &
\multicolumn{3}{c}{5.16\footnotemark[1]} &
\multicolumn{3}{c}{from $-20$ to $-30$\footnotemark[2]} \\
\end{tabular}
\end{ruledtabular}
\footnotetext[1]{Reference \onlinecite{KoskenmakiHPCRE78}.}
\footnotetext[2]{References \onlinecite{AmadonPRL06,DecrempsPRL11}.}
\end{table*}

\begin{table*}
\caption{\label{table2}Spin magnetic moment in the unit cell
$M_{\text{spin,cell}}$ (in $\mu_{\text{B}}$), orbital magnetic moment of the
$4f$-electrons $M_{\text{orb},4f}$ (in $\mu_{\text{B}}$), and number of
$4f$-electrons $n_{4f}$ inside the atomic sphere
($R_{\text{MT}}^{\text{Ce}}=2.2$ bohr). The results for LDA, PBE, LDA+$U$,
and PBE+$U$ include SOC effects.}
\begin{ruledtabular}
\begin{tabular}{lcccccccccc}
Method                             & $M_{\text{spin,cell}}^{\text{FM1}}$ & $M_{\text{spin,cell}}^{\text{FM2}}$ & $M_{\text{spin,cell}}^{\text{FM3}}$ & $M_{\text{orb},4f}^{\text{FM1}}$ & $M_{\text{orb},4f}^{\text{FM2}}$ & $M_{\text{orb},4f}^{\text{FM3}}$ & $n_{4f}^{\text{NM}}$ & $n_{4f}^{\text{FM1}}$ & $n_{4f}^{\text{FM2}}$ & $n_{4f}^{\text{FM3}}$ \\
\hline
LDA                                &                                     &                                     &                                     &                                  &                                  &                                  & 0.95                 &                       &                       &                       \\
PBE                                &                                     &                                     &                                     &                                  &                                  &                                  & 0.91                 &                       &                       &                       \\
LDA+$U$ ($U=2.7$ eV)               &                                     &                                     & 1.1                                 &                                  &                                  & $-1.6$                           & 0.78                 &                       &                       & 0.95                  \\
LDA+$U$ ($U=4.3$ eV)               &                                     &                                     & 1.2                                 &                                  &                                  & $-1.6$                           & 0.69                 &                       &                       & 0.97                  \\
PBE+$U$ ($U=1.5$ eV)               & 1.2                                 & 1.4                                 & 1.2                                 & $-0.5$                           & $-0.6$                           & $-1.5$                           & 0.82                 & 0.95                  & 0.99                  & 0.93                  \\
PBE+$U$ ($U=4.3$ eV)               & 1.2                                 & 1.3                                 & 1.2                                 & $-0.5$                           & $-0.5$                           & $-1.8$                           & 0.65                 & 0.96                  & 0.97                  & 0.96                  \\
YS-PBEh ($\alpha_{\text{x}}=0.08$) & 1.2                                 &                                     &                                     &                                  &                                  &                                  & 0.81                 & 0.93                  &                       &                       \\
YS-PBEh ($\alpha_{\text{x}}=0.25$) & 1.1                                 &                                     &                                     &                                  &                                  &                                  & 0.61                 & 0.94                  &                       &                       \\
\end{tabular}
\end{ruledtabular}
\end{table*}

\begin{figure}
\includegraphics[scale=0.8]{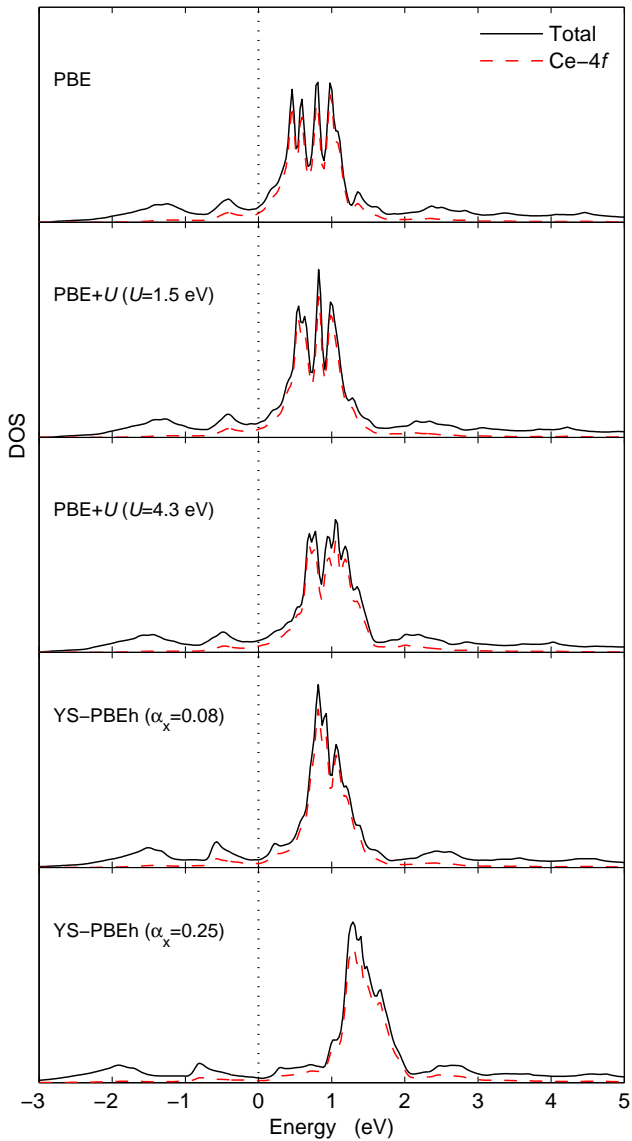}
\caption{\label{fig3}(Color online) Total and $4f$ one-spin density of states
for the NM phase of cerium. The results for PBE and PBE+$U$ include SOC effects.
The Fermi energy is set at $E=0$ eV.}
\end{figure}

\begin{figure}
\includegraphics[scale=0.8]{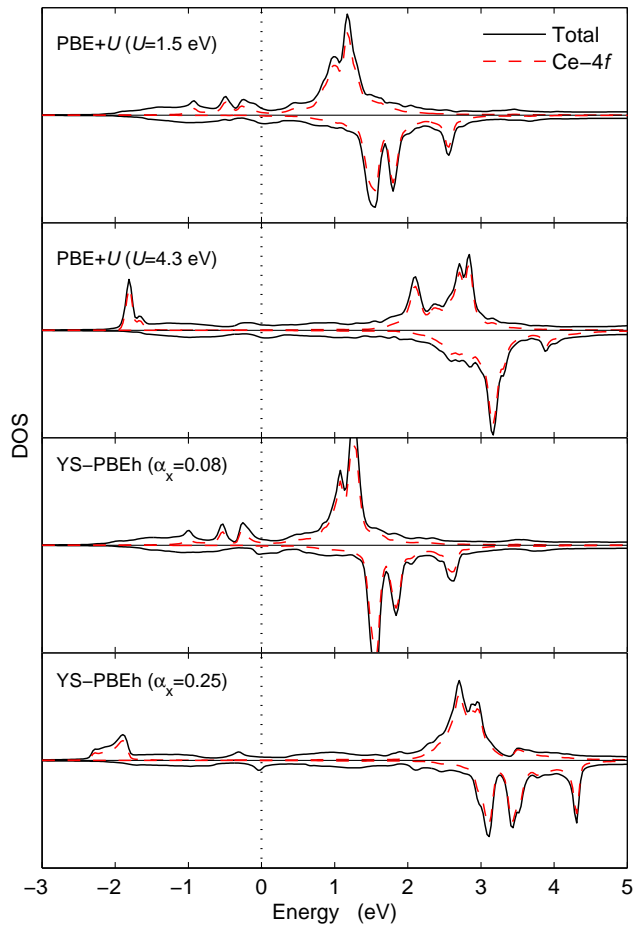}
\caption{\label{fig4}(Color online) Total and $4f$ spin-majority (upwards)
and spin-minority (downwards) density of states for the FM
phase of cerium. The FM state is FM2 for PBE+$U$ with $U=4.3$ eV and FM1 in
the other cases. The results for PBE+$U$ include SOC effects.
The Fermi energy is set at $E=0$ eV.}
\end{figure}

\begin{figure}
\includegraphics[scale=0.8]{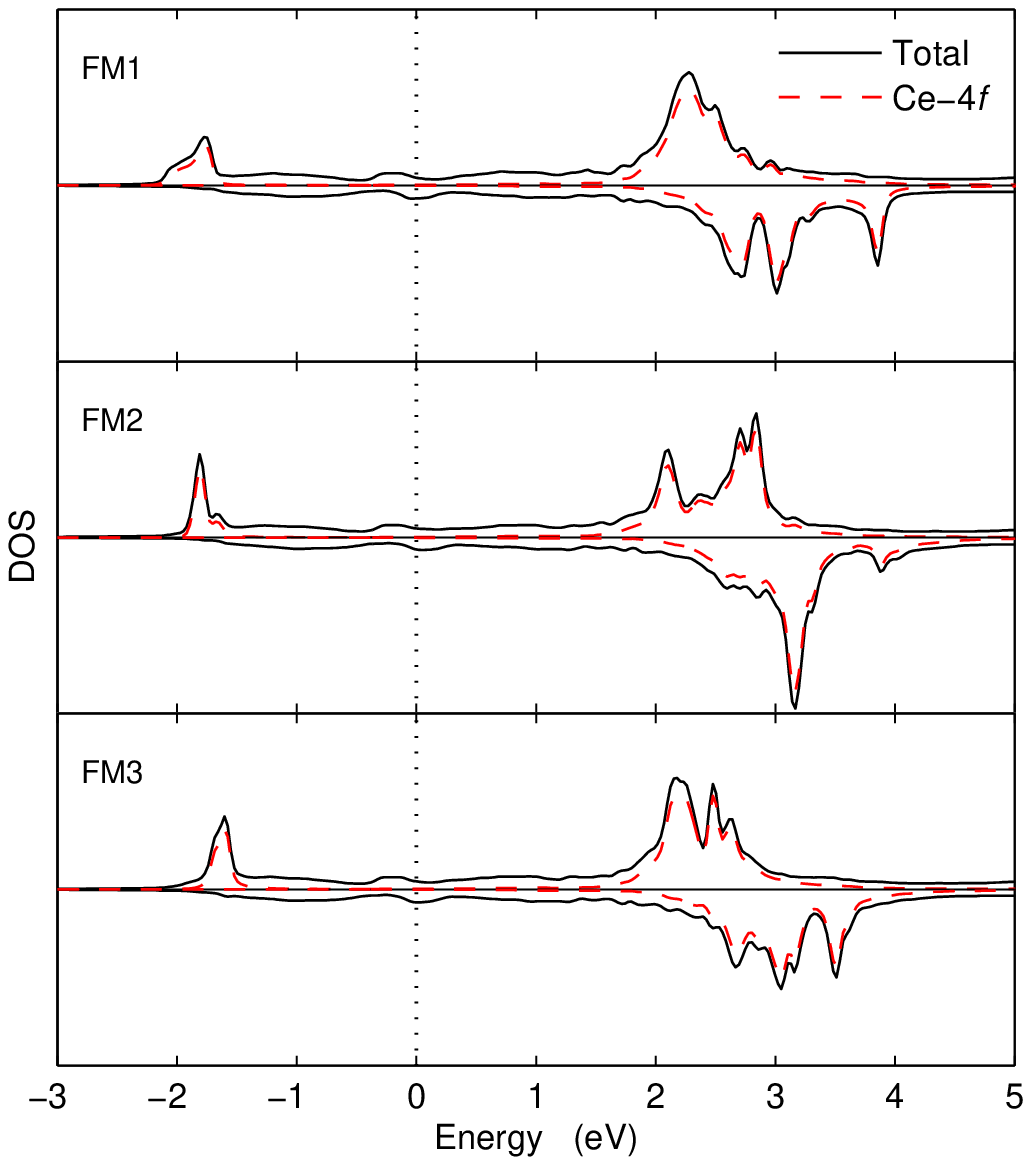}
\caption{\label{fig5}(Color online) Total and $4f$ spin-majority (upwards)
and spin-minority (downwards) density of states for the three
FM phases of cerium as obtained from PBE+$U$ ($U=4.3$ eV) with SOC.
The Fermi energy is set at $E=0$ eV.}
\end{figure}

In the 0$-$0.25 range of fraction $\alpha_{\text{x}}$ of HF exchange in YS-PBEh,
two values are of interest and the corresponding results are shown in
Fig. \ref{fig2} and Table \ref{table1}. $\alpha_{\text{x}}=0.08$ is a value
which leads to rather similar results as PBE+$U$ with $U=1.5$ eV
and therefore in fair agreement with experiment, except for the lattice constant
of the $\alpha$ phase which is still underestimated by about 0.1 \AA.
The energy difference between the NM and FM1 phases is $\Delta E=-88$ meV,
which seems to be slightly too large in magnitude but still reasonable at a
qualitative level, while the transition pressure is at $P_{t}=-1.7$ GPa.
The value $\alpha_{\text{x}}=0.25$ is a non-empirical value which
was deduced from perturbation theory arguments
(see Ref. \onlinecite{PerdewJCP96}). The results obtained with
$\alpha_{\text{x}}=0.25$ show a rather large overestimation of 0.15 \AA~
for the lattice constant of the FM1 phase and the wrong stability
ordering ($\Delta E=576$ meV) of the two phases (similar to PBE+$U$ with $U=4.3$ eV).
Note that, these results with $\alpha_{\text{x}}=0.25$ agree
with the HSE06 results of Casadei \textit{et al},\cite{CasadeiPRL12}
but with the difference that in their $\alpha$ phase a non-zero magnetic moment
of about 0.2 $\mu_{\text{B}}$ is obtained. It is also worth mentioning
that their RPA results show the right
stability ordering, however the lattice constants are too small, in particular
for the $\alpha$ phase (underestimations of 0.4 and 0.1 \AA~for the
$\alpha$ and $\gamma$ phases, respectively). 

In Ref. \onlinecite{LanataPRL13}, the results of calculations obtained with
the LDA+GA and LDA+DMFT methods at the temperature $T=0$ K were reported.
It was concluded that the phase transition (indicated by a change of sign
in the bulk modulus) can be observed only if SOC is included in the
calculations. As already mentioned in Sec. \ref{details}, SOC
has very little influence on the results of our calculations.

We finish this section by mentioning that the FM2 and FM3 solutions (only
reachable if cubic symmetry is broken) lead to structural distortion (not
included in the results shown in Table \ref{table1} and Fig. \ref{fig2}).
By considering tetragonal distortion along
the $z$-direction (i.e., $c$-axis),
we calculated a ratio $c/a$ of about 0.97 and a lowering of the total energy
(with respect to the cubic unit cell) of the order of 2 meV, which is one order
of magnitude smaller than the relative stability of the $\alpha$ and $\gamma$
phases. The NM and FM1 solutions (including SOC) lead to no distortion.
However, we mention that the observed distortions for the FM2 and FM3 solutions
are artifacts in the sense that Ce is paramagnetic (and not ferromagnetic as
in our work) such that the randomness of the orientations of the local moments
would cancel any structural distortion.

\subsection{\label{electronicstructure}Electronic structure}

Photoemission spectroscopy (PES)\cite{WuilloudPRB83,WieliczkaPRB84,WeschkePRB91}
and inverse PES\cite{WuilloudPRB83} experiments have shown that lower and
upper Hubbard bands are present in the $\alpha$ and $\gamma$ phases, which
indicates that the $4f$-electrons are strongly correlated in both phases.
The lower Hubbard band is situated at about $-2.2$ and $-2$ eV below
the Fermi energy in the $\alpha$ and $\gamma$ phases, respectively, while
the upper Hubbard band is at 4 eV above the Fermi energy in both
phases. In addition, in the $\alpha$ phase a quasiparticle peak (Kondo resonance)
is observed at the Fermi energy.

Figures \ref{fig3} and \ref{fig4} show the calculated DOS obtained with the
PBE-based methods of the NM and FM phases, respectively.
In the NM case, we can see that the
occupied part of the Ce-$4f$ partial DOS is rather flat and extends from
$-1$ to 0 eV below the Fermi energy, which is in disagreement with experiment,
and actually, there is no clearly separated lower and upper Hubbard bands.
In general, the features of the NM DOS are pretty similar among all considered
functionals since the value of $U$ or $\alpha_{\text{x}}$ seems to have a
moderate influence on the position of the Ce-$4f$ partial DOS.
Note that since in the NM phase, the main part of the $4f$ DOS is situated
just above the Fermi energy, it is tempting to assign it to
the observed quasiparticle peak, however, it is questionable wether such
a feature which originates from many-body effects can be described by
one-electron methods.

Figure \ref{fig4} (FM DOS) shows that for the small value of $U$ (1.5 eV) or
$\alpha_{\text{x}}$ (0.08), the occupied Ce-$4f$ DOS (the lower Hubbard band)
is relatively flat and in the range $[-1,0]$ eV (as for the NM phase), while
for larger $U$ (4.3 eV) or $\alpha_{\text{x}}$ (0.25), the Ce-$4f$ DOS is
sharper and is shifted down at $-2$ eV below the Fermi energy, which is in
good agreement with experiment. As previously shown in
Refs. \onlinecite{ShickJESRP01,AmadonPRB08}, LDA+$U$ with $U-J=5.4$ eV puts
the lower Hubbard band at $-2.5$ eV in the $\gamma$ phase.
We can see that larger values of $U$ or $\alpha_{\text{x}}$ also lead
to a more correct position of the upper Hubbard band.

With PBE+$U$, the small sensitivity of the Ce-$4f$ DOS to $U$ in the NM
case can be simply explained by the fact that the diagonal terms of the occupation
matrix [see Eq. (\ref{ocNM})] are more or less of equal magnitude, such that the
shift due to the orbital-dependent $U$-potential
[see Eq. (25) in Ref. \onlinecite{YlvisakerPRB09}] is similar
wether the $4f$ orbital is below or above the Fermi energy.
In the FM case, the occupied $4f$-orbital corresponds to one (or two)
particular value of $m$ [Eqs. (\ref{ocFM1})-(\ref{ocFM3})], which
allows the $U$-potential to shift [by $\sim(U-J)/2$] the lower and upper Hubbard
bands in opposite directions. With the hybrid functional YS-PBEh, whose
potential is also orbital-dependent, a similar mechanism occurs.
As already observed for the lattice constants and relative energies, the
PBE+$U$ and YS-PBEh results are pretty similar for the electronic
structure, too.

In order to show the effect of orbital occupation on the DOS, we show in
Fig. \ref{fig5} the DOS of the three FM states, all obtained from
the same method (PBE+$U$ with $U=4.3$ eV and including SOC). As we can see,
the symmetry of the occupied $4f$ orbital has overall little influence on
the position of the center of mass of the lower and upper Hubbard bands.

Overall, it seems that it is not possible to reproduce all important features
seen in the (inverse) PES experiments. More specifically, none of the
one-electron methods that we have considered is able to yield a spectrum for
the $\alpha$ phase showing (simultaneously) the Hubbard bands and the
quasiparticle peak.
In the recent study of Sakuma \textit{et al}. (Ref. \onlinecite{SakumaPRB12}),
it has been shown that the non-self consistent $GW$ method
($GW$ on top of LDA) can also not reproduce correctly all main features
of the experimental spectrum (no lower Hubbard band at $-2$ eV and
presence of a quasiparticle peak also in the $\gamma$ phase).
By now, only LDA+DMFT, which properly takes into account many-body effects,
is able to yield good agreement with experiment.
\cite{ZolflPRL01,HeldPRL01,SakaiJPSJ05,HaulePRL05,McMahanPRB05,AmadonPRL06}
It is worth to mention that our calculated DOS in the NM phase looks
rather similar to the LDA+DMFT spectrum obtained at very small
lattice constants (see Refs. \onlinecite{HeldPRL01,McMahanPRB05}).

From neutron inelastic-scattering experiment,\cite{MuraniPRB93} it was
inferred that the number of $4f$-electrons $n_{4f}$ is smaller in the
$\alpha$ phase than in the $\gamma$ phase by $0.2\pm0.1$. The values
of $n_{4f}$ inside the atomic sphere ($R_{\text{MT}}^{\text{Ce}}=2.2$ bohr)
shown in Table \ref{table2} reproduce this trend, albeit the difference
$n_{4f}^{\text{FM}}-n_{4f}^{\text{NM}}$ seems too be at the limit of being
too large ($0.33$) for PBE+$U$ with $U=4.3$ eV and YS-PBEh with
$\alpha_{\text{x}}=0.25$. On the side of LDA+DMFT,
some discrepancies among the various studies were obtained.
For instance, while in Ref. \onlinecite{AmadonPRL06},
$n_{4f}$ was calculated to be larger in the $\alpha$ phase for
temperatures ranging from 400 to 1600 K (a monotonous increase
upon compression is obtained), the opposite trend
(and with a non-monotonous behavior of $n_{4f}$) was obtained
in Ref. \onlinecite{McMahanPRB03} for temperatures below roughly 1000 K.
Note, however, that the value of $n_{4f}$, and possibly the trend in the
variation due to volume change, depends on the basis set and more particularly
on the size of the atomic sphere.

The results for the magnetic moments are shown in Table \ref{table2}.
We can see that for the spin moment in the unit cell $M_{\text{spin,cell}}$,
PBE+$U$ and YS-PBEh lead to a value of 1.1$-$1.2 $\mu_{\text{B}}$ for the FM1
solution, whatever is the value of $U$ or $\alpha_{\text{x}}$.
The values are similar for FM3, while they are slightly larger by
0.1$-$0.2 $\mu_{\text{B}}$ for FM2. The contribution to $M_{\text{spin,cell}}$
coming from inside the atomic sphere amounts to 0.9$-$1.0 $\mu_{\text{B}}$ and
the rest comes from the interstitial region. SOC induces an orbital moment and
for the $4f$ shell ($M_{\text{orb},4f}$), we can see that
for FM1 and FM2 the values are similar ($-0.5$ $\mu_{\text{B}}$), while
they are much larger for FM3 since this solution corresponds
mainly to an electron in the $Y_{3}^{-2}$ orbital.

\section{\label{summary}Summary}

The purpose of this work has been to study with KS and mixed
KS/HF methods the FM and NM phases of elemental cerium. Several types
of functionals were considered and for two of them, DFT+$U$ and
YS-PBEh, it was possible to get a minimum in the total-energy curve for
both phases without imposing any constraint on the spin symmetry.
The parameters $U$ and $\alpha_{\text{x}}$ in these functionals
were varied in order to examine their influence on the properties.

In order to compare our results with experiment, we have supposed
that our NM and FM solutions correspond to the $\alpha$ and
$\gamma$ phases that were observed experimentally. We have shown that
the correct stability ordering of the $\alpha$ and $\gamma$ phases
can be obtained only for small values of $U$ or $\alpha_{\text{x}}$.
On the other hand, the electronic structure is better reproduced with
larger values of $U$ or $\alpha_{\text{x}}$, but none of the considered
methods is able to give an overall correct description of the electronic
structure. In particular, up to now only the many-body LDA+DMFT method has been
able to reproduce the Hubbard bands and quasiparticle peak in the $\alpha$
phase. In this respect, it would be very interesting to know how
would perform the HF+RPA method when applied self-consistently
as a one-electron method.\cite{KotaniJMMM98}

\begin{acknowledgments}

This work was supported by the project SFB-F41 (ViCoM) of the Austrian Science Fund.

\end{acknowledgments}

\appendix

\section{\label{occupation}Occupation Matrix}

In this appendix, the majority-spin $4f$ occupation matrices $n_{mm'}$ of the FM and
NM solutions (at their respective equilibrium volume) obtained from
PBE+$U$ with $U=4.3$ eV and including SOC are given.

FM1:
\begin{equation}
\begin{pmatrix}
0.00 &   0.00  & 0.00 & 0.00 & 0.00 &   0.00  & 0.00 \\
0.00 &   0.59  & 0.00 & 0.00 & 0.00 & $-0.44$ & 0.00 \\
0.00 &   0.00  & 0.00 & 0.00 & 0.00 &   0.00  & 0.00 \\
0.00 &   0.00  & 0.00 & 0.01 & 0.00 &   0.00  & 0.00 \\
0.00 &   0.00  & 0.00 & 0.00 & 0.00 &   0.00  & 0.00 \\
0.00 & $-0.44$ & 0.00 & 0.00 & 0.00 &   0.33  & 0.00 \\
0.00 &   0.00  & 0.00 & 0.00 & 0.00 &   0.00  & 0.00 \\
\end{pmatrix}
\label{ocFM1}
\end{equation}

FM2:
\begin{equation}
\begin{pmatrix}
0.00 & 0.00 & 0.00 & 0.00 & 0.00 & 0.00 & 0.00 \\
0.00 & 0.59 & 0.00 & 0.00 & 0.00 & 0.44 & 0.00 \\
0.00 & 0.00 & 0.00 & 0.00 & 0.00 & 0.00 & 0.00 \\
0.00 & 0.00 & 0.00 & 0.00 & 0.00 & 0.00 & 0.00 \\
0.00 & 0.00 & 0.00 & 0.00 & 0.00 & 0.00 & 0.00 \\
0.00 & 0.44 & 0.00 & 0.00 & 0.00 & 0.34 & 0.00 \\
0.00 & 0.00 & 0.00 & 0.00 & 0.00 & 0.00 & 0.00 \\
\end{pmatrix}
\label{ocFM2}
\end{equation}

FM3:
\begin{equation}
\begin{pmatrix}
0.00 &   0.00  & 0.00 & 0.00 & 0.00 &   0.00  & 0.00 \\
0.00 &   0.92  & 0.00 & 0.00 & 0.00 & $-0.02$ & 0.00 \\
0.00 &   0.00  & 0.00 & 0.00 & 0.00 &   0.00  & 0.00 \\
0.00 &   0.00  & 0.00 & 0.01 & 0.00 &   0.00  & 0.00 \\
0.00 &   0.00  & 0.00 & 0.00 & 0.00 &   0.00  & 0.00 \\
0.00 & $-0.02$ & 0.00 & 0.00 & 0.00 &   0.00  & 0.00 \\
0.00 &   0.00  & 0.01 & 0.00 & 0.00 &   0.00  & 0.00 \\
\end{pmatrix}
\label{ocFM3}
\end{equation}

NM:
\begin{equation}
\begin{pmatrix}
0.05 &   0.00  & 0.00 & 0.00 & 0.03 &   0.00  & 0.00 \\
0.00 &   0.08  & 0.00 & 0.00 & 0.00 & $-0.06$ & 0.00 \\
0.00 &   0.00  & 0.03 & 0.00 & 0.00 &   0.00  & 0.02 \\
0.00 &   0.00  & 0.00 & 0.06 & 0.00 &   0.00  & 0.00 \\
0.03 &   0.00  & 0.00 & 0.00 & 0.03 &   0.00  & 0.00 \\
0.00 & $-0.06$ & 0.00 & 0.00 & 0.00 &   0.06  & 0.00 \\
0.00 &   0.00  & 0.02 & 0.00 & 0.00 &   0.00  & 0.03 \\
\end{pmatrix}
\label{ocNM}
\end{equation}

\end{document}